\documentclass[pre]{revtex4}

\usepackage{graphicx,amssymb,color}
\usepackage[dvipsnames]{xcolor}

\newcommand{\beq}{\begin{equation}}
\newcommand{\eeq}{\end{equation}}

\begin{document}
\title{ Supporting Material for: 
Capillary-like Fluctuations of a Solid-Liquid Interface in a Non-Cohesive Granular System}
\author{Li-Hua Luu, Gustavo Castillo, Nicol\'as Mujica and Rodrigo Soto}
 
\affiliation{%
Departamento de F\'{\i}sica, Facultad de Ciencias F\'{\i}sicas y Matem\'aticas Universidad de Chile,\\
Avenida Blanco Encalada 2008, Santiago, Chile}%
\date{\today}

\begin{abstract}

In this supplementary paper we present some details on the solid-liquid interface detection, the deduction of the non-equilibrium free energy, the analysis of the granular temperature and energy per mode, a validation of the small slope approximation, a description of the Langevin dynamics, and the error analysis.

\end{abstract}

\maketitle

\subsection{Experimental determination of solid-liquid interface}
\label{sec_expint}

In the quasi-2D geometry the solid phase consists of two square interlaced layers instead of the single hexagonal layer that is characteristic of 2D systems \cite{urbach}. These interlaced square lattices, when projected in 2D, result also in a square lattice. If particles in the solid phase are permanently in contact such that grains are close packed, then the unit  cell per layer should have an area $d\times  d$, implying that the projected lattice should have a Voronoi area equal to $d^2/2$. 

Two local parameters can be used to analyze the system. For each particle we can compute its Voronoi cell area, $A_v$, which is inversely proportional to the local density, and also the 4-fold local order parameter, $Q_4$,  that is defined as \cite{Castillo}
\begin{equation}
Q_4^j = \frac{1}{N_{j}} \sum_{s = 1}^{N_{j}} e^{4i\alpha_{s}^j}.
\end{equation}
Here, $N_j$ is the number of nearest neighbors of particle $j$ and $\alpha_s^j$ is the angle between the neighbor $s$ of particle $j$ and the $x$ axis.  For a particle in a square lattice, $|Q_4^j | = 1$ and the complex phase measures  the square lattice orientation respect to the $x$ axis.

\begin{figure}[t!]
\begin{center}
\includegraphics[width=12cm]{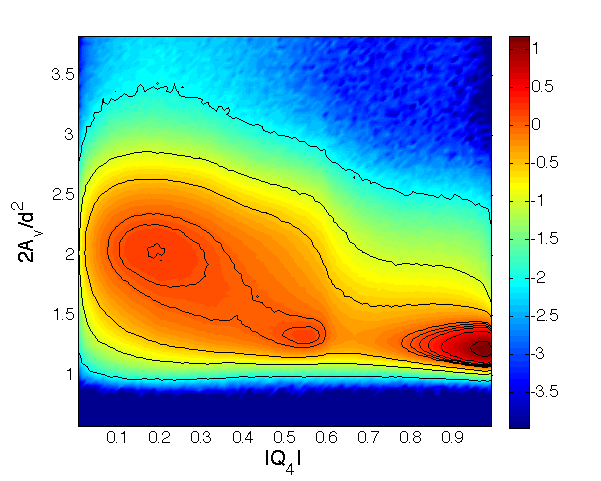}
\caption{Probability density function of normalized Voronoi areas $2A_v/d^2$ and local order parameter $|Q_4|$ obtained from 3000 images (about $2.8\times 10^6$ pairs of $A_v$ and $|Q_4|$ values), for which the system is indeed phase separated. The criterion for solid-liquid identification relies on the valley obtained around $|Q_4|=0.7$ that is almost independent of $A_v$. The color code indicated in the vertical bar is the logarithm (base 10) of the PDF. Contour plots are shown for $\log_{10}[{\rm PDF(|Q_4|,A_v)}] = -2$, $-1$, $-0.5$, $0$, $0.1$, $0.2$, $0.3$, $0.5$, $1$.}
\label{fig1}
\end{center}
\end{figure}

\begin{figure}[t!]
\begin{center}
\includegraphics[width=8cm]{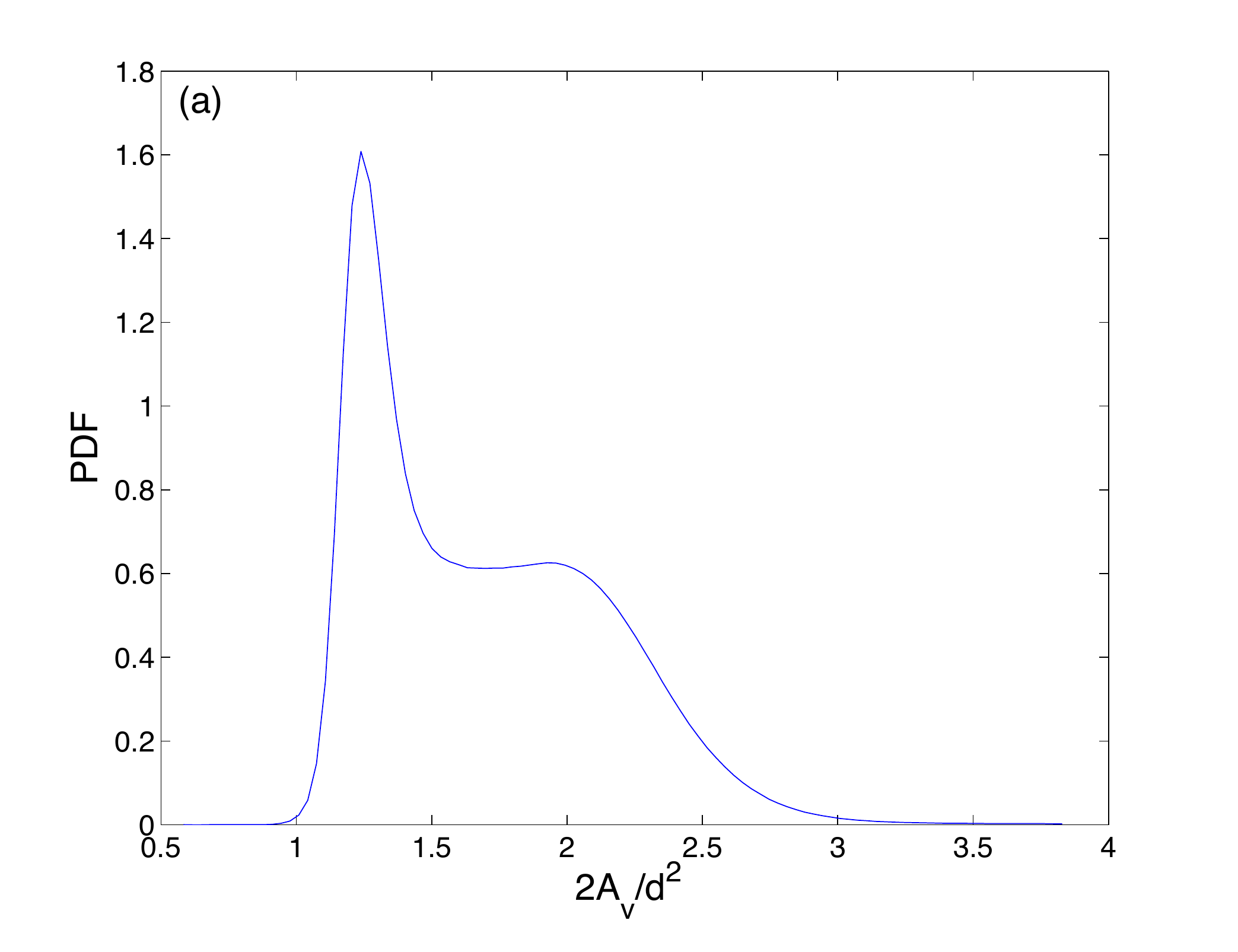}
\includegraphics[width=8cm]{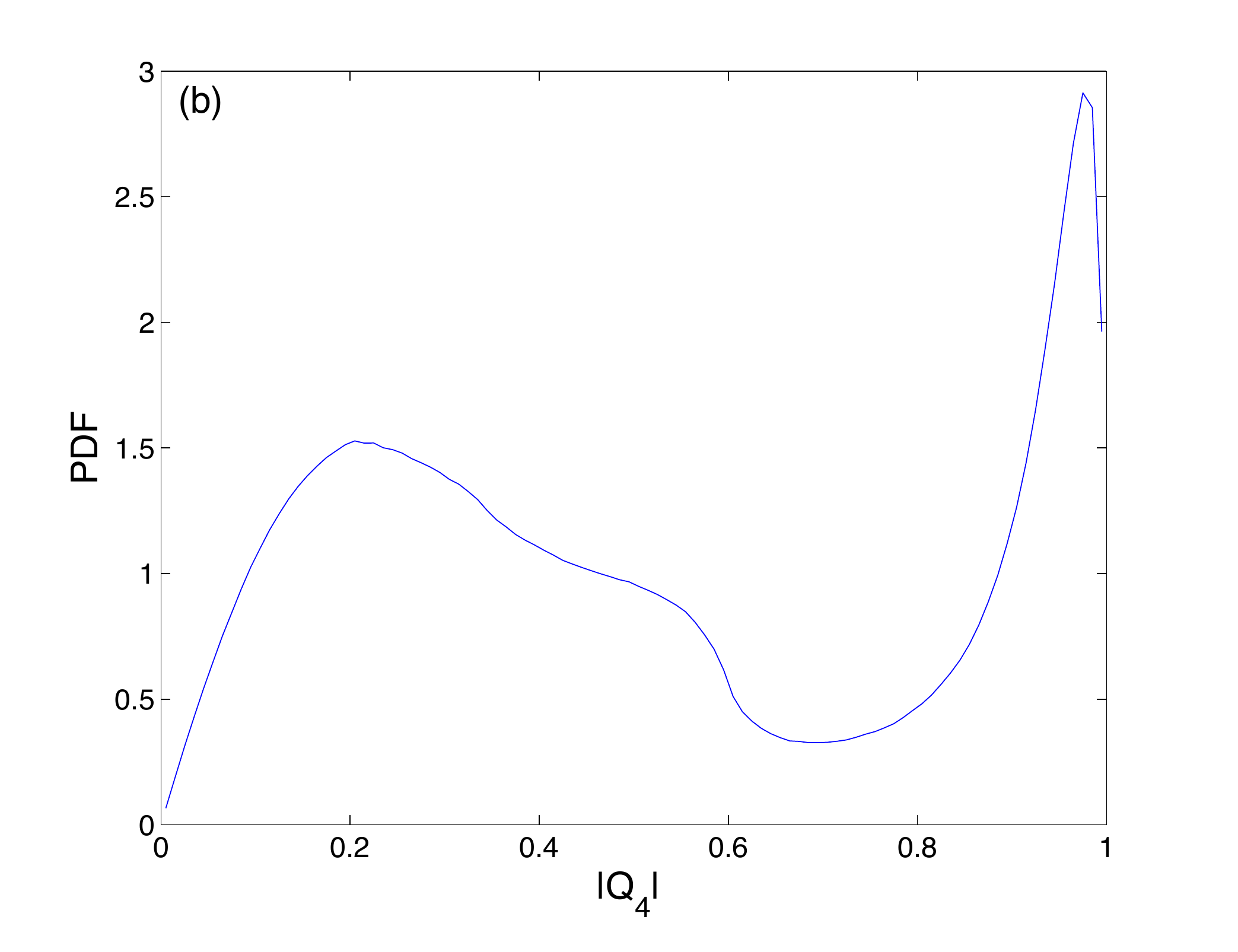}
\caption{Projected probability density functions of normalized Voronoi areas $2A_v/d^2$ (a) and local order parameter $|Q_4|$ (b) obtained from 3000 images. The minimum obtained for $|Q_4|=0.7$ is chosen as criterion for solid-liquid identification.}
\label{fig2}
\end{center}
\end{figure}

Fig. \ref{fig1} presents the probability distribution function (PDF) of normalized Voronoi areas $2A_v/d^2$ and local order parameter $|Q_4|$ obtained from 3000 images, for which the system is indeed phase separated. The Voronoi area, $A_v$, is normalized by $d^2/2$, the corresponding Voronoi area for closed packed two square interlaced layers. Fig. \ref{fig2} shows the marginal PDF, defined as 
\begin{eqnarray}
{\rm PDF}(A_v) &=& \int{\rm PDF}(|Q_4|,A_v) dA_v,\\
{\rm PDF}(|Q_4|) &=& \int{\rm PDF}(|Q_4|,A_v) d|Q_4|. 
\end{eqnarray}
From the joint and marginal PDFs (Fig. \ref{fig1} and Fig. \ref{fig2} respectiveley) we can state that there is clearly a superposition of three distributions: a wide distribution related to the liquid phase, around $|Q_4|\approx 0.2$ and $2A_v/d^2 \approx 2$; another more localized and stronger peak related to the solid phase, close to $|Q_4|=1$ and $2A_v/d^2 \approx 1.3$; and a third small peak of more ordered particles in a dense phase, with $|Q_4|\approx0.55$ and $2A_v/d^2 \approx 1.4$, which correspond to small density and order fluctuations in the system. In general, the normalized Voronoi area of particles in the solid phase is larger than 1, implying that particles are slightly separated (in average). A separation distance of $0.15d$ is enough to shift this peak from $1$ to $1.3$. 

For classifying particles in the solid or liquid phase three simple criteria can be established. The first one is to use only $A_v$, defining a critical value $A_v^c$ ($2A_v^c/d^2 \approx 1.5$) such that particles with $A_v$ lower (larger) than this value are labelled as solid (liquid) particles. A second criterion is to use only local order. If $|Q_4| \geqslant Q_4^c$ ($|Q_4| < Q_4^c$), then the particle is in the solid (liquid) phase. A third possibility is to use both quantities, such that if $A_v \leqslant A_v^c$ and $|Q_4| \geqslant Q_4^c$ are both satisfied, then the particle is considered to be in the solid phase. We have verified that our results are not sensitive to which condition we use (see section \ref{demspectra}), and for simplicity we choose the second condition, with  $|Q_4^c|=0.7$. However, the first criterion also detects particles that are less ordered but dense, adding a thin liquid-like layer, with intermediate order, around the solid cluster.

We note also that in Ref. \cite{Castillo} we concluded that $Q_4$ is a more appropriate order parameter to characterize the liquid to solid transition, compared to the local density. Indeed the correlations of $Q_4$ showed critical divergencies at the critical point, while those of the local density were insensitive to the transition.

Figure 1 of the main text shows a typical example of the solid-liquid interface detection. In our experiment, for each image particles that are in a solid or liquid phase are labeled using the previously defined criterion. Also for each image, particles belonging to the same cluster are identified, as well as the largest cluster in each image. This cluster of course corresponds to the largest, most stable, solid cluster that is present when the driving acceleration exceeds the critical value. Then, for each image the center of mass of particles that belong to the largest solid cluster is determined. Using all the centers of masses from a stack of images, the global center of mass is determined. 

The next step then corresponds to detect those particles that are at the solid-liquid interface of the largest solid cluster for each image. This is done by an angle coarse-graining procedure. From the position of the global center of mass, the position of the solid particle that is the most distant from this center of mass within an angle $\Delta \theta$ is determined. This is done from $\theta_1 = 0^\circ$ until $\theta_N = 360^\circ-\Delta \theta$, following $\theta_i = \theta_1 + (i-1)\Delta \theta$. Thus, for each $\theta_i$, there is a pair of coordinates $(x_i^b, y_i^b)$ of the particle that is in the largest solid cluster and that is the most distant to the global center of mass, thus at the solid-liquid boundary. For simplicity, we describe each of these  particles by the coordinates $(\theta_i^b,r_i^b)$, which are polar coordinates of each particle respect to the reference system fixed at the global center of mass. The final step is to interpolate these coordinates to a set where the angles are equally spaced, such that $\theta_i^{\rm int} = \theta_1 + (i-1)\Delta \theta/2$, and we end with the set of coordinates $(\theta_i^{\rm int},r_i^{\rm int})$. It is this set of coordinates that is used for the Fourier analysis presented in this study. 

Figure 1 of the main text also shows that in general the interface is rather smooth, except for some ``jumps". These jumps are artificial, as they arise because we force the interface to be described by a single-evaluated function $r_i^{\rm int}= r_i^{\rm int}(\theta_i^{\rm int})$, which is not always the case. Because of the adopted procedure for the interface determination, we necessarily end with a few of these jumps for each image. However, theses jumps do not contribute to the low wavenumber modes that turn out to be relevant for this study. This is explicitly shown in section \ref{sect_small_slope}, where we discuss the validity of the small slope approximation.

\subsection{Non-equilibrium free energy}
The solid-liquid interface non-equilibrium free energy of a curved surface is written in two dimensions as
\begin{equation}
E =  \gamma \int \limits_{0}^{2\pi} \sqrt{R^2 + (\partial_\theta R)^2} d\theta.
\label{ecn_E}
\end{equation}
The total non-equilibrium free energy has an additional term, because if we minimize the solid-liquid interface non-equilibrium free energy, the absolute minimum is at $R=0$, that is, no droplet. In principle, this is solved by adding a Lagrange multiplier that fixes (in average) the total mass (or equivalently, the area). Then, the total non-equilibrium free energy should be
\begin{equation}
E =  \gamma \int \limits_{0}^{2\pi} \sqrt{R^2 + (\partial_\theta R)^2} d\theta - \mu \int \limits_{0}^{2\pi} \frac{R^2}{2} d\theta.
\label{ecn_E2}
\end{equation}
where $\mu$ is the Lagrange multiplier, which apart from a proportional factor is equivalent to the effective chemical potential. The negative sign is arbitrary and it is there to further simplify the notation.

The equilibrium droplet is obtained minimizing the non-equilibrium free energy. First, non-uniformities in the radius increase the non-equilibrium free energy, so we assume a circular shape $R=R_0$. Then the non-equilibrium free energy is
\begin{equation}
E=2\pi \gamma R_0 - \mu \pi R_0^2. \label{eq_ER0}
\end{equation} 
To extreme $E$ we take the derivative and equal it to zero. The solution is 
\beq
R_0=\frac{\gamma}{\mu} ,
\eeq
or equivalently $\mu=\gamma/R_0$. Note that this last expression shows that we have correctly chosen the sign for $\mu$ in (\ref{ecn_E2}).

Now, we linearize about $R_0$. That is, $R(\theta) =R_0 + \epsilon \delta R(\theta)$ and keep up to quadratic terms in $\epsilon$ in the energy. Doing the expansion we obtain
\beq
E =  \gamma \int \limits_{0}^{2\pi} \left[ R_0 + \epsilon  \delta R + \epsilon^2 \frac{(\partial_{\theta} \delta R)^2}{2R_0}  \right]  d\theta 
- \frac{\mu}{2} \int \limits_{0}^{2\pi} \left[ R_0^2 + 2 \epsilon R_0 \delta R + \epsilon^2 \delta R^2 \right]  d\theta.
\eeq
Recalling that $\mu=\gamma/R_0$ the linear terms cancel (as it always should happen when doing an expansion about an equilibrium state) and the final result is
\beq
E = \pi\gamma R_0 +\frac{\epsilon^2 \gamma}{2R_0} \int \limits_{0}^{2\pi} \left[(\partial_{\theta} \delta R)^2 - \delta R^2  \right]  d\theta 
\eeq
This expression does not describe our system. Indeed, the non-equilibrium free energy increases due to deformations of the interface but is decreases when changing globally the radius.  That is, $R_0$ is unstable under changes of the radius.
In fact, this inconsistency could have been predicted when looking at the expression (\ref{eq_ER0}): it is clear that $R_0$ is a maximum and, therefore, un unstable equilibrium.

\subsubsection{The origin of the problem}
The problem originates in the election of the thermodynamic ensemble. Indeed, when using $\mu$ we moved to the grand canonical ensemble in which the number of particles is not fixed. What we have found is what is called the Critical Nucleus in the {\em Homogeneous Nucleation Theory}. This is the size we have to overcome to create a  droplet that, after reaching that size, will grow indefinitely. Droplets smaller than this radius will shrink again due to the energy cost of the free surface. The growth without bound is possible because we are in the grand canonical ensemble where we have fixed $\mu$ and, therefore, there are always available particles to change phase.

\subsubsection{The solution}
The solution consists on changing the ensemble to one in which we fix the total number of particles. This is a complicated ensemble to make calculations because we have to take into account the physical fact that when particles are moved to one phase to another, the {\em supersaturation} changes and therefore the chemical potential is dynamically adjusted. This is precisely what happens when, working in the canonical ensemble, we prepare the system with a density between the liquid and solid densities. Clusters of the solid phase will be created, decreasing the density of the remainder part until it reaches the liquid density.

To model this solution, we can write in the case of a circular droplet
\begin{equation}
E=2\pi \gamma R_0 - \mu(R_0) \pi R_0^2. \label{eq2_ER0}
\end{equation} 
If $\mu$ is a decreasing function of $R_0$, a second equilibrium appears (this time stable) at a larger value of $R_0$. This is the final equilibrium  radius of the droplet.

\subsection{Full model}
Equivalently of using a chemical potential that depends on the radius, we modify (\ref{ecn_E2}) to have a non-linear dependence on the total mass of the cluster, expression that will be easier to manipulate
\begin{equation}
E =  \gamma \int \limits_{0}^{2\pi} \sqrt{R^2 + (\partial_\theta R)^2} d\theta - f\left(\int \limits_{0}^{2\pi} \frac{R^2}{2} d\theta \right),
\label{ecn_E3}
\end{equation}
where $f$ is a nonlinear function. For a cluster of area much smaller than the total system area, $f(x)=\mu x$ as in (\ref{ecn_E2}).

Again, an expansion up to quadratic terms is done obtaining
\begin{eqnarray}
E &=&   \gamma \int \limits_{0}^{2\pi} \left[ R_0 + \epsilon  \delta R + \epsilon^2 \frac{(\partial_{\theta} \delta R)^2}{2R_0}  \right]  d\theta \nonumber \\
&&- \left\{ f\left(\pi R_0^2 \right) + \epsilon f'\left(\pi R_0^2\right) R_0 I_1
+\frac{\epsilon^2}{2} \left[ f'\left(\pi R_0^2\right) I_2 + f''\left(\pi R_0^2\right)R_0^2 I_3\right] \right\} ,
\label{ecn_E4}
\end{eqnarray}
where 
\begin{eqnarray}
I_1 &=& \int \delta R d\theta\\
I_2 &=& \int (\delta R)^2 d\theta\\
I_3 &=& \left(\int \delta R d\theta\right)^2 .
\end{eqnarray}

The equilibrium radius is such that the linear terms should cancel (as to have a global minimum). This condition gives
\beq
f'(\pi R_0^2) R_0 = \gamma ,
\eeq
that once substituted in (\ref{ecn_E4}) gives
\begin{eqnarray}
E &=&  E_0 + \frac{\epsilon^2 \gamma}{2 R_0} \int \limits_{0}^{2\pi} \left[(\partial_{\theta} \delta R)^2 - \delta R^2  \right]  d\theta 
-\frac{\epsilon^2}{2} f''(\pi R_0^2) R_0^2  \left(\int \limits_{0}^{2\pi}  \delta R d\theta\right)^2 .
\label{ecn_E5}
\end{eqnarray}

Fourier-transforming the last expression gives
\beq
E=E_0 + \frac{\epsilon^2 \pi \gamma}{R_0} \sum_m |\widetilde{\delta R}_m|^2 (m^2-1) 
 - \frac{\epsilon^2}{2} f''(\pi R_0^2) (2\pi R_0)^2 |\widetilde{\delta R}_0|^2,  \label{energypermode}
\eeq
where
\beq
\widetilde{\delta R}_m = \frac{1}{2\pi} \int \delta R\, e^{-i m \theta}\,d\theta 
\eeq

In the expression (\ref{energypermode}) all terms can be correctly interpreted. 
First, each mode contributes to the non-equilibrium free energy with a curvature dependence $m^2$, proportional to the surface tension. However, the modes $m=\pm 1$ do not contribute. This is indeed correct because when a circle is slightly (linear in $\epsilon$) perturbed in the modes $m=\pm1$, it is only translated but not deformed (this is related to the absence of dipolar perturbation of a circle or sphere) and can be directly checked as follows. Take the radius and do a $m=1$ deformation, for example $R=R_0(1+\epsilon \cos(\theta))$, then it is simple to check that $(x-\epsilon R_0)^2+y^2=R_0^2$, that is the circle is just translated. As there is  translational symmetry these modes should not contribute to the energy. Finally, the last term corresponds to the increase of energy when the mass of the cluster is changed. To assure that the equilibrium is stable, the second derivative of $f$ should be negative and its absolute value larger than the first term. Note, however, that its value is otherwise completely independent of $\gamma$, that is the mass fluctuations are independent of the shape fluctuations. In summary, the non-equilibrium free energy about the equilibrium droplet can be written as
\beq
E=E_0 +  \frac{\pi \lambda}{R_0} |\widetilde{\delta R}_0|^2 + \frac{\pi \gamma}{R_0} \sum_{|m|\geq 2} |\widetilde{\delta R}_m|^2 (m^2-1) ,
\eeq
where the formal expansion parameter $\epsilon$ has been suppressed and $\lambda$ is a new parameter, which has the same units as $\gamma$. In experiments, the translational symmetry is not perfect and the cluster has a tendency to remain in the center of the box. Then, the non-equilibrium free energy should be modified to
\beq
E=E_0 + \underbrace{ \frac{\pi \lambda}{R_0}  |\widetilde{\delta R}_0|^2}_{\mbox{\tiny change in size}} + \underbrace{  \frac{\pi \nu}{R_0} \left( |\widetilde{\delta R}_1|^2 + |\widetilde{\delta R}_{-1}|^2 \right) }_{\mbox{\tiny change in position}} + \underbrace{\frac{\pi \gamma}{R_0} \sum_{|m|\geq 2} |\widetilde{\delta R}_m|^2 (m^2-1) }_{\mbox{\tiny change in form}} 
\label{ecn_completa_E}
\eeq
with $\nu$ a new parameter, with the same units as $\gamma$ and $\lambda$. This expression is used to derive the equilibrium power spectrum and the time correlation functions. 

The parameters $\lambda$ and $\nu$ have physical origins that are different from the surface tension parameter $\gamma$, so their three values do not have to be directly related. $\lambda$ corresponds to a measure of solid cluster's size changes, either by expansion and contraction (at a fixed number of particles) or by the condensation and evaporation of particles in and out of the solid cluster. $\nu$ corresponds to a measure of how far from the ideal condition is the current experimental realization. Ideally, $\nu = 0$, which implies that this mode (translation) is a neutral mode, equivalent to the spatial average for a flat interface (limit $k\rightarrow 0$). If $\nu=0$, the $m=1$ mode would realize a simple random walk, as discussed in section \ref{sect_dynamic_corr}. In fact, we show below that this is not the case. We speculate that the origin of $\nu>0$ is related to particle's vertical dynamics, and thus depends on surface roughness, friction, and top and bottom local wall parallelism. In particular, it depends on the vertical dynamics of the whole solid cluster, which is still largely unknown.

\subsection{Static power spectrum}

Starting from (\ref{ecn_completa_E}), for equilibrium systems the equipartition theorem from statistical mechanics is invoked. Thus, each
normal mode, when configurationally averaged (in our case when time averaged), contributes $k_B T /2$ of energy to the system. Thus, the power spectrum is
\begin{eqnarray}
\langle |\widetilde {\delta R}_0|^2 \rangle &=&  \frac{k_B T R_0}{2 \pi \lambda},\\
\langle |\widetilde {\delta R}_{\pm 1}|^2 \rangle &=&  \frac{k_B T R_0}{2 \pi \nu}, \\
\langle |\widetilde {\delta R}_m|^2 \rangle &=&  \frac{k_B T R_0}{2 \pi \gamma (m^2-1)},
\label{ecn_powerspect}
\end{eqnarray}
where the last line is valid for $|m|\geqslant 2$. 

In general, the equipartition theorem is used for the total kinetic and potential energies (if any) by defining the number of active modes $n$. Thus, in average 
\beq
\langle K \rangle = \langle U \rangle = n \,\frac{1}{2} k_B T,
\eeq
where $K$ and $U$ are the total kinetic and potential energies respectively. The idea is that active modes contribute with $k_BT/2$ to the total energy, whereas non-active modes do not contribute. The number of active modes $n$ depends on the particular system, by the number of spatial dimensions and by the particular atom or molecule interactions (intramolecular and intermolecular). The most simple example is an ideal monoatomic gas in $D_s$ dimensions, with no potential energy. In this case, $n = D_s N$, where $N$ is the number of atoms.

In our case, we consider that the granular system has an effective ``granular" temperature $T_{\rm eff}$. Although this quantity is commonly referred as a temperature it has units of energy. A simple version is to consider it equivalent to $k_B T$. But, our non-equilibrium system shows that this effective thermal energy, or granular temperature, has to be defined carefully, as well as the number of active modes. In our setup, the system is isotropic in the horizontal plane, but the horizontal thermal energy is different from the vertical one. For the following analysis we focus our attention to the projected 2D dynamics, which is what can be analyzed experimentally. In fact, figure 2b of the paper shows that equipartition can be applied for the horizontal kinetic energy, up to a given wavenumber (from $m=2$ to $m\approx 30$). 

We now focus on the solid-liquid interface, as a subsystem of the complete granular system. We recall that the protocol used identifies one interface particle for each angle interval $\Delta\theta =2^\circ$. Therefore, the interface subsystem consists always in exactly $N_p=180$ particles.
The interface total potential energy is given by equation (\ref{ecn_completa_E}). We then consider equipartition for the total kinetic and potential energies, 
\beq
\langle K \rangle = \langle E \rangle = n \,\frac{1}{2} T_{\rm eff},
\eeq
where $n$ is the number of active modes and $T_{\rm eff}$ is the effective granular temperature. The important issue is that we can measure $\langle K \rangle$ and express it as a sum of normal modes. Indeed, 
\beq
\langle K \rangle = \frac{1}{2} \sum \limits_{i = 1}^{N_p} m_p \langle  \vec v_i^2 \rangle =  \frac{1}{2} N_p m_p \langle  \vec v^2 \rangle
\eeq
where $\langle  \vec v^2 \rangle =  \langle   v_x^2 \rangle +  \langle   v_y^2 \rangle$ is the velocity variance. Here, we must stress that because $ \vec v$ is function of both $\theta$ and $t$, the average $\langle\,\rangle$ is a double average, over angles and over time.  Both velocity components can be expressed by Fourier series:
\beq
 v_x = \sum \limits_{m=-\infty}^\infty {{\tilde v}_m}^x e^{i m \theta}, \hspace{2cm}  v_y = \sum \limits_{m=-\infty}^\infty {{\tilde v}_m}^y e^{i m \theta},
\eeq
where ${{\tilde v}_m}^x$ and ${{\tilde v}_m}^y$ are the Fourier components. In practice, because $\Delta \theta = 2^\circ$, then the summation is done for a finite number of modes, from $m= -m^*$ to $m= m^*$, with $m^* = N_p/2-1 = 89$. Thus, 
\beq
\langle K \rangle =  \frac{1}{2} N_p m_p \sum \limits_{m=-m^*}^{m^*} \langle |{{\tilde v}_m}^x|^2 + |{{\tilde v}_m}^y|^2 \rangle, 
\eeq
which allows to define the kinetic granular energy per mode, 
\beq
\langle K_m \rangle =  \frac{1}{2} N_p m_p \langle |{{\tilde v}_m}^x|^2 + |{{\tilde v}_m}^y|^2 \rangle.
\eeq
For the last two expressions, the angle average has already been performed, and $\langle\,\rangle$ is a configurational (time) average. This is the quantity plotted in figure 2b of the paper, which shows equipartition from $m=2$ to $m\approx 30$ (corresponding to wavelengths $\approx 74d$ to $\approx 5d$). In fact, an average between $m=2$ and $m=30$ gives $\langle K_m \rangle \equiv K_{\rm eq}= 4.82\pm 0.04$ nJ. The two lowest modes, which correspond to global radius fluctuations and center of mass translations,  have higher energy: $\langle K_0 \rangle = 8.0^{+4.7}_{-4.6}$ nJ and $\langle K_1 \rangle = 5.9\pm{2.8}$ nJ. Their errors are slightly asymmetric due to their asymmetric (non-Gaussian) distribution functions, which are detailed in the last section of this document. 

Furthermore, the identity 
\beq
\langle K \rangle =  \frac{1}{2} N_p m_p \langle  \vec v^2 \rangle = \frac{1}{2} N_p m_p \sum \limits_{m=-m^*}^{m^*} \langle |{{\tilde v}_m}^x|^2 + {{\tilde v}_m}^y|^2 \rangle,
\eeq
can be verified numerically. Indeed, from the particle velocity distributions (in $x$ and $y$) and their variances, we obtain $\langle K \rangle \approx 0.80$ $\mu$J, whereas from the velocity Fourier components we obtain $\langle K \rangle \approx 0.82$ $\mu$J.

Next, as we measure for each mode $\langle K_m \rangle$, we can apply $\langle K_m \rangle = \langle \delta E_m \rangle $, where  $\langle \delta E_m \rangle$ corresponds to each mode contribution to 
\begin{equation}
\langle \delta E \rangle= { \frac{\pi \lambda}{R_0}  \langle |\widetilde{\delta R}_0|^2 \rangle} +  \frac{\pi \nu}{R_0} \left( \langle |\widetilde{\delta R}_1|^2\rangle 
+ \langle |\widetilde{\delta R}_{-1}|^2 \rangle \right) + \frac{\pi \gamma}{R_0} \sum_{|m|\geq 2} \langle |\widetilde{\delta R}_m|^2 \rangle (m^2-1).
\label{ecn_completa_E2}
\end{equation}
Thus, the final expressions for the static power spectrum for our system are
\begin{eqnarray}
\langle |\widetilde {\delta R}_0|^2 \rangle &=&  \frac{\langle K_0 \rangle  R_0}{\pi \lambda},\\
\langle |\widetilde {\delta R}_{\pm 1}|^2 \rangle &=&  \frac{\langle K_1 \rangle  R_0}{\pi \nu}, \\
\langle |\widetilde {\delta R}_{|m|\geqslant 2}|^2 \rangle &=&  \frac{\langle K_{|m|\geqslant 2} \rangle  R_0}{\pi \gamma (m^2-1)}.
\label{ecn_powerspect2}
\end{eqnarray}

\subsection{Capillary-like spectrum for different interface detection conditions}
\label{demspectra}

In section A we have presented the interface detection procedure, including the three different criteria used for the selection of particles in the solid cluster. In the main text we also discuss the validity of the capillary-like spectrum respect to the choice of the coarse-graining angle $\Delta \theta$. 

\begin{figure}[t!]
\begin{center}
\includegraphics[width=8cm]{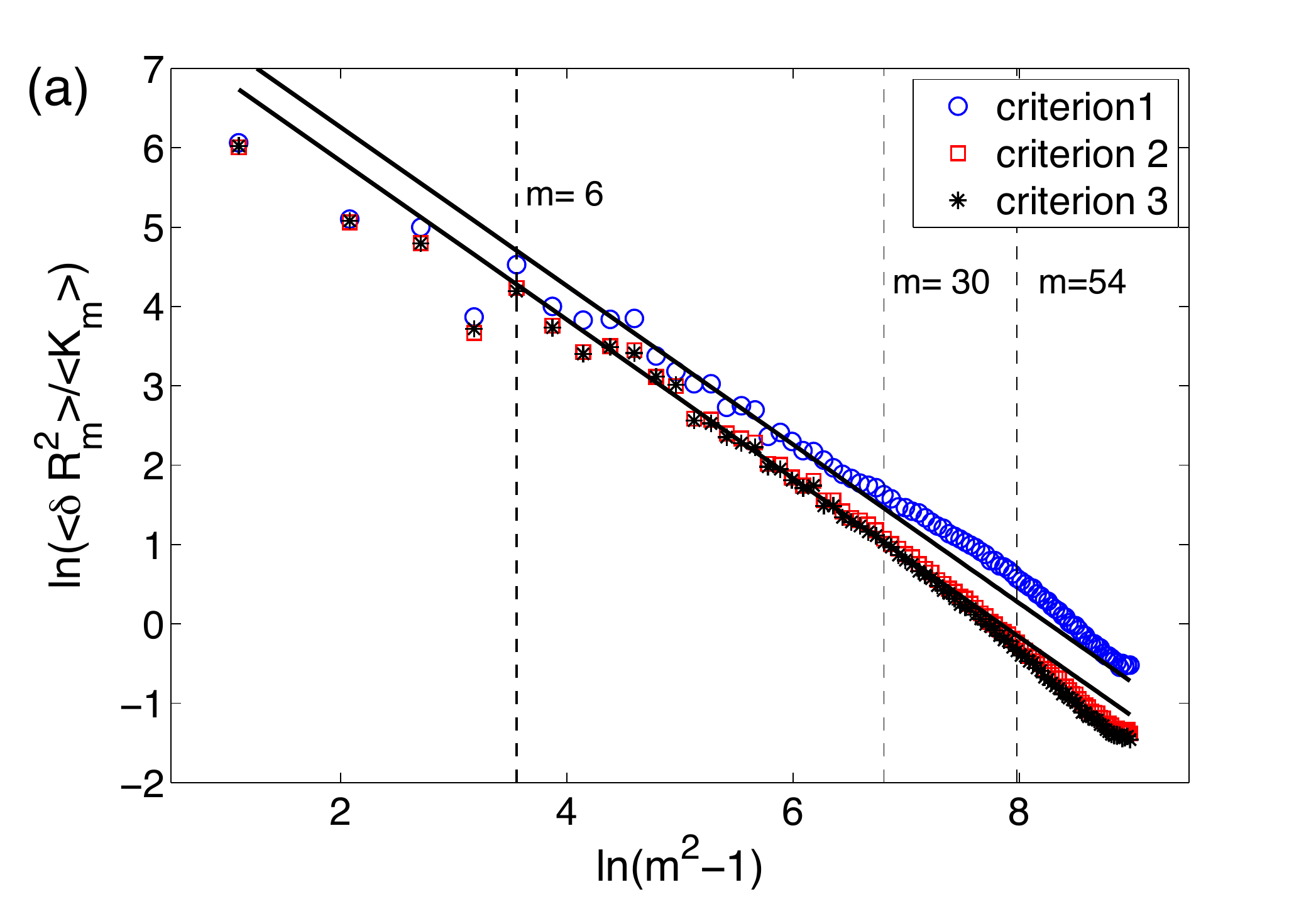}\\
\includegraphics[width=8cm]{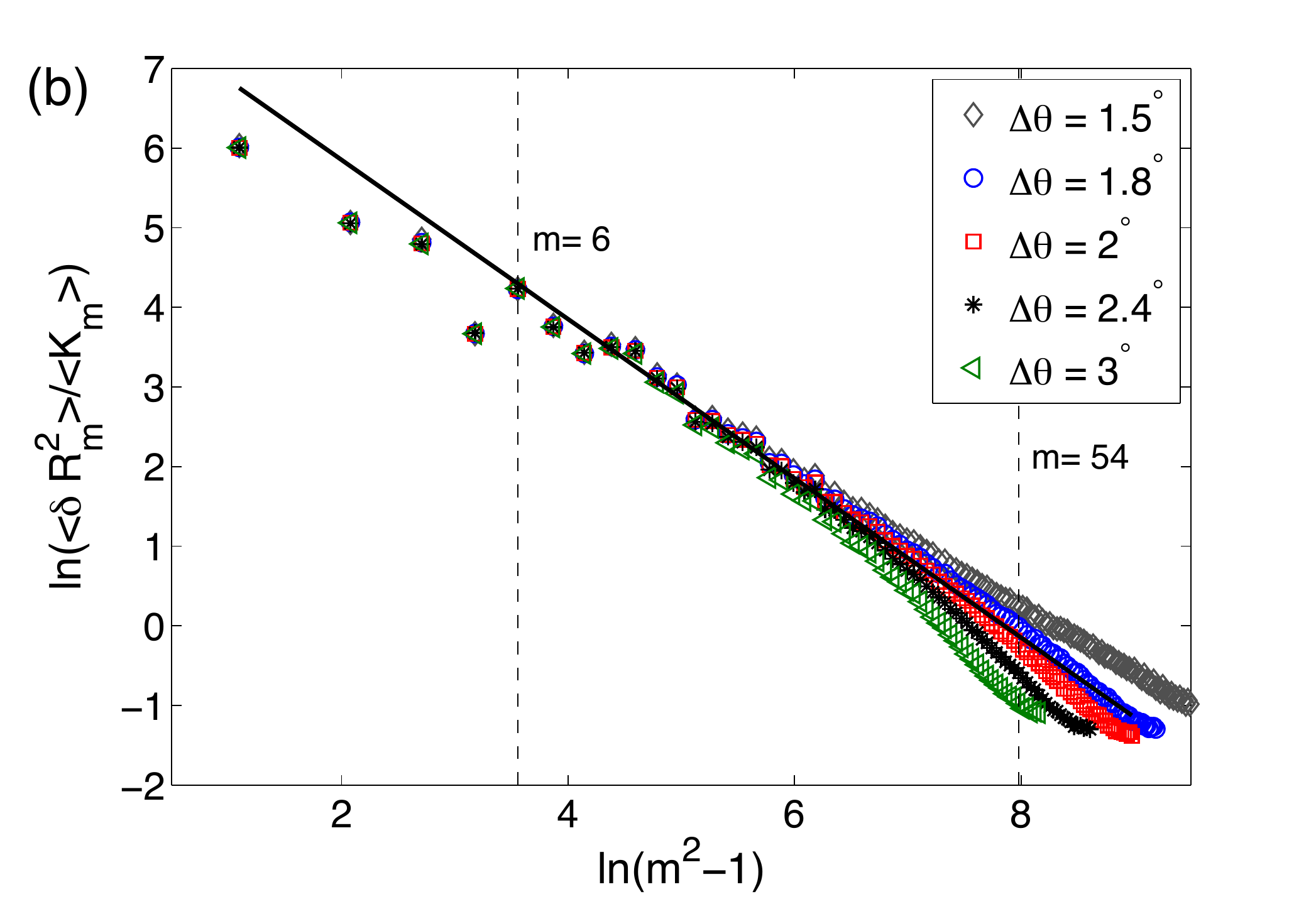}
\includegraphics[width=8cm]{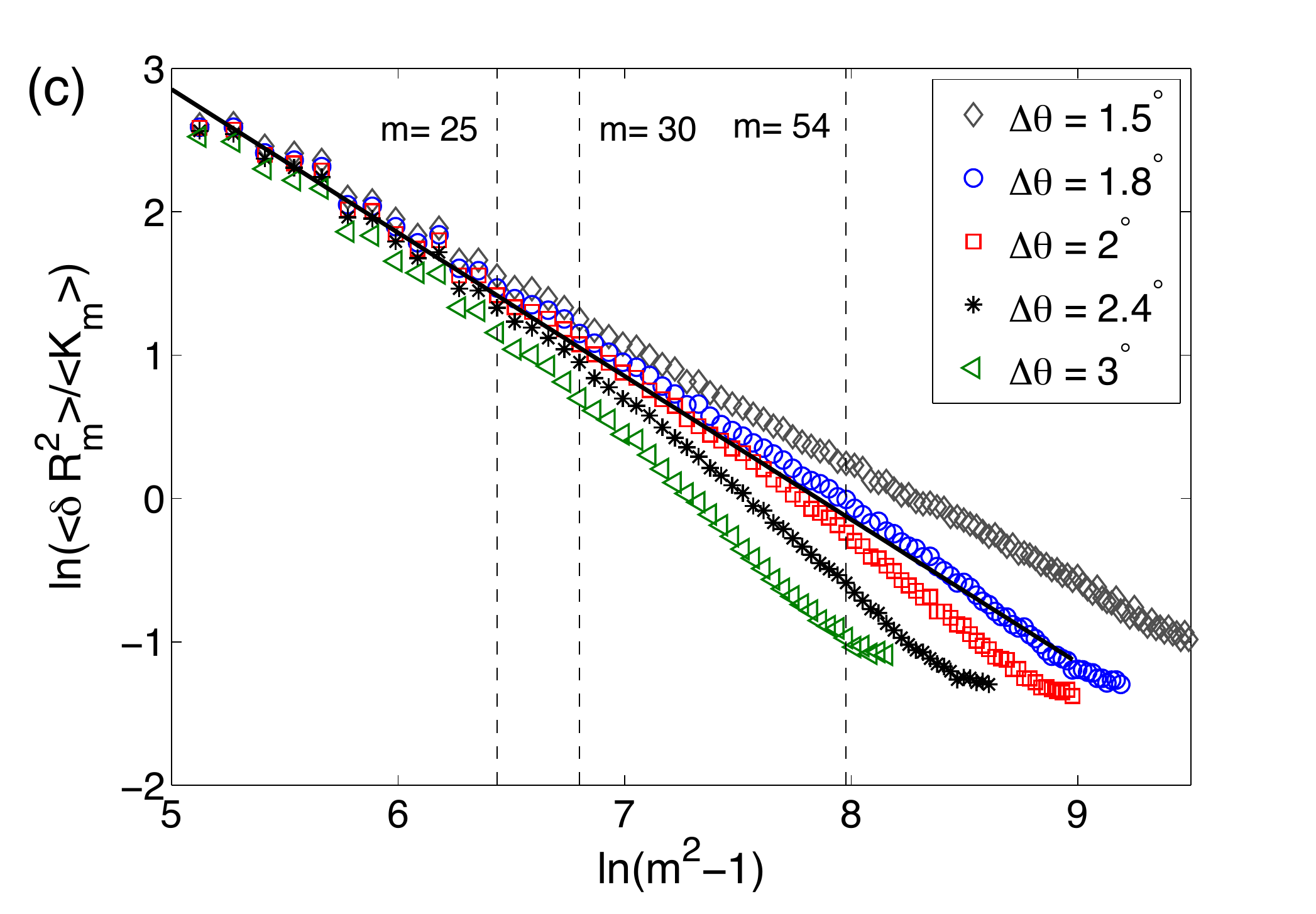}
\caption{(a) Fluctuation spectra $\langle |\widetilde {\delta R}_{m}|^2 \rangle/\langle K_m \rangle$ versus $m^2-1$  for $\Delta\theta = 2^\circ$ and the three  cluster selection criteria presented in section \ref{sec_expint}. The vertical dashed lines show the limit values used for the fits: $m=6...30$ for criterion 1, and $m=6...54$ for criteria 2 and 3. (b)  Fluctuation spectra $\langle |\widetilde {\delta R}_{m}|^2 \rangle/\langle K_m \rangle$ versus $m^2-1$ for several $\Delta \theta$ using criterion 2. For smaller $\Delta\theta$, noise contaminates the spectrum at higher $m$ values. For larger $\Delta \theta$, the coarse-graining procedure acts as a filter at larger $m$. (c) Same as (b) for a smaller range of $m$, putting in evidence the discrepancies that appear for larger $m$.}
\label{figspectrum}
\end{center}
\end{figure}

In figure \ref{figspectrum} we present several fluctuation spectra $\langle |\widetilde {\delta R}_{m}|^2 \rangle/\langle K_m \rangle$ versus $m^2-1$ using the three different criteria and several coarse-graining angles. The $1/(m^2-1)$ dependence is observed for all cases, with variations on the range of mode numbers $m$ for which is it valid, depending on the specific details of the detection procedure. For $\Delta \theta = 2^\circ$ (Fig. \ref{figspectrum}a) we obtain that for the first criterion the fitted surface tension results $\gamma = 2.0\pm 0.1$~$\mu$N. For the second and third criteria, within errors their result are equal,  $\gamma = 2.9\pm 0.1$~$\mu$N. Figs. \ref{figspectrum}b and \ref{figspectrum}c present spectra for the second criterion with different coarse-graining angles. 

These results allow us to conclude that the description of the fluctuations by the capillary theory is very robust with respect to the details of the interface detection procedure. However, the numerical parameters of the theory (surface tension and mobility) are more sensible to the different criteria. In particular, the fact that $\gamma$ is lower for criterion 1 can be explained recalling that the detection procedure adds a thin liquid-like skin to the solid cluster, as mentioned in section \ref{sec_expint}. This layer is composed by particles with intermediate order ($|Q_4|\approx 0.55$) but with small Voronoi area ($2A_v/d^2 \approx 1.4$), thus with higher density.

\subsection{About the small slope approximation}
\label{sect_small_slope}

\begin{figure}[ht!]
\begin{center}
\includegraphics[width=14cm]{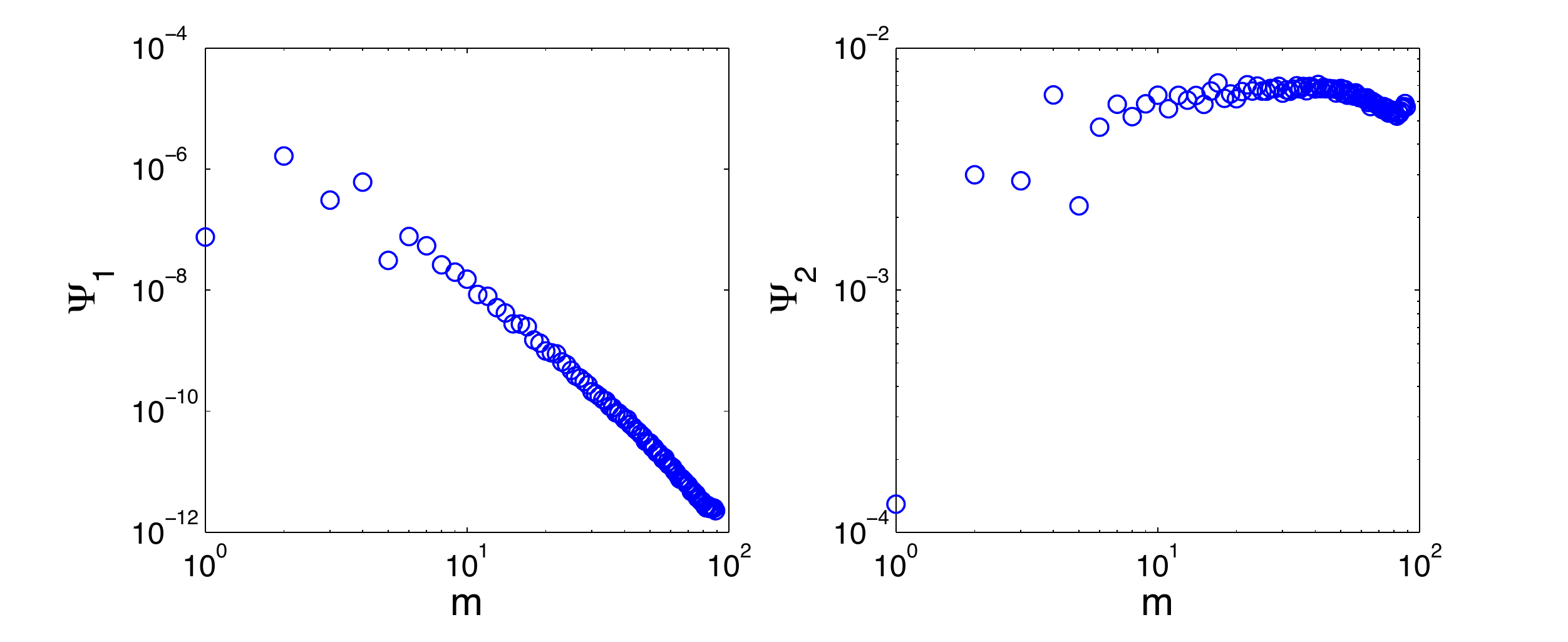}
\caption{Relative intensity of the correction terms, $\Psi_1 = \langle |\widetilde{\delta R_m}|^4 \rangle/(8 R_0^4)$ (left) and  $\Psi_2 = m^2 \langle | \widetilde{\delta R_m}|^2 \rangle/(4 R_0^2)$ (right), to the quadratic form of the non-equilibrium free energy in (\ref{eq.corrtermsslope}). Small values imply that the small slope approximation to the non-equilibrium free energy is valid and the quadratic form can be used.} 
\label{fig.smallslope}
\end{center}
\end{figure}

The small slope approximation is used explicitly in the Taylor expansion of $\sqrt{R^2 + (\partial_\theta R)^2}$. In particular, the last two terms in 
\bigskip
\bigskip
\begin{eqnarray}
\sqrt{R^2 + (\partial_\theta R)^2} &=& \sqrt{(R_0+\delta R)^2 + (\partial_\theta \delta R)^2} \\
&\approx& R_0 \left [ 1 + \frac{\delta R}{R_0} + \frac{1}{2}\left ( \frac{\partial_\theta \delta R}{R_0} \right ) ^2 -  \frac{1}{8}\left ( \frac{\delta R}{R_0} \right ) ^4 - \frac{1}{8}\left ( \frac{\partial_\theta \delta R}{R_0} \right ) ^4 + h.o.t \right ], \label{eq.corrtermsslope}
\end{eqnarray}
are neglected, as well as the higher order terms (h.o.t) that are not shown. As discussed previously, the contribution of the linear term vanishes in the non-equilibrium free energy. Thus, we are left to show that the following conditions are indeed fulfilled by our experimentally determined solid-liquid interfaces: 
\begin{eqnarray}
 \frac{1}{8}\left ( \frac{\delta R}{R_0} \right ) ^4 &\ll& 1,\\
  \frac{1}{8}\left ( \frac{\partial_\theta \delta R}{R_0} \right ) ^4 &\ll& \frac{1}{2}\left ( \frac{\partial_\theta \delta R}{R_0} \right ) ^2.
\end{eqnarray}
These terms contribute to the non-equilibrium free energy through the angle integral. Thus, by expanding $\delta R$ in Fourier series, the above expressions become equivalent to show these conditions are satisfied: 
\begin{eqnarray}
\Psi_1 \equiv \frac{\langle |\widetilde{\delta R_m}|^4 \rangle}{8 R_0^4} \ &\ll& 1,\\ 
\Psi_2 \equiv m^2 \frac{\langle | \widetilde{\delta R_m}|^2 \rangle}{4 R_0^2}  &\ll& 1.
\end{eqnarray}
Here, the brackets $\langle\,\, \rangle$ represent an ensemble average, which in practice is done through time average of data from many images. These conditions are indeed satisfied, as shown in Fig.~\ref{fig.smallslope}. The first parameter satisfies  $\Psi_1 <1.7 \times 10^{-6}$ for all $m$, whereas the second one, $\Psi_2 <8 \times 10^{-3}$ for all~$m$.

\subsection{Dynamic correlation function: determination of effective mobility}
\label{sect_dynamic_corr}

We now focus on the mobility parameter $M$ and the interface dynamic correlation function. For an interface of flat geometry, with local height $h(x,t)$, the standard procedure is to consider the interface local velocity $V = \partial h/\partial t = M \gamma \kappa$,
where $\kappa$ is the interface's curvature ($\kappa = \partial_{xx}h$ for a flat interface in 2D). The Fourier transform of this equation leads to
\beq
\frac{\partial A_k}{\partial t} = - M \gamma k^2 A_k,
\label{ecn_Ak}
\eeq
where we have used the Fourier representation
\beq
h(x,t) =  \sum_{k} A_k(t) e^{ikx}.
\eeq
When a noise term is included, Langevin equations are obtained, in real and Fourier space respectively:
\beq
\frac{\partial h(x,t)}{\partial t} = M \left (\gamma \frac{\partial^2 h(x,t)}{\partial x^2} + \eta(x,t) \right ),  \quad\quad \frac{\partial A_k}{\partial t} = M (- \gamma k^2 A_k + \eta_k(t)), 
\eeq
where 
\beq
\eta(x,t) =  \sum_{k} \eta_k(t) e^{ikx}
\eeq
is modeled as a white noise term, which is delta correlated
\begin{eqnarray}
\langle \eta(x,t) \rangle &=& 0,\\
\langle \eta(x,t)\eta(x',t') \rangle &=& C \delta(x-x')\delta(t-t'),\\
\langle \eta_k(t) \rangle &=& 0,\\
\langle \eta_k(t)\eta^*_k(t') \rangle &=& C \delta_{k,k'}\delta(t-t').
\end{eqnarray}
In 2D, if we consider an interface of length $L$ between two semi-infinite phases, then the non-equilibrium free energy is 
\beq
E = \gamma \int_{0}^{L} \sqrt{1 + (\partial_{x} h)^2} dx \approx \gamma L + \frac{\gamma L}{2} \sum_{k} k^2 |A_k|^2.
\eeq
It is straight forward to show that Eqn. (\ref{ecn_Ak}) can be obtained by the general form
\beq
\frac{\partial A_k}{\partial t} = -\frac{M}{L} \frac{\delta E}{\delta A_k}, 
\eeq
where $\delta E/\delta A_k$ is the functional derivative of $E$. 

Applying the same formalism to our case, and including noise, from Eqn. (\ref{ecn_completa_E}) we obtain the following Langevin equations
\begin{eqnarray}
\frac{\partial  \widetilde{\delta R}_0}{\partial t} &=& M \left ( - \frac{\lambda}{R_0^2}  \widetilde{\delta R}_0 + \eta_0(t)  \right ),\\
\frac{\partial  \widetilde{\delta R}_{\pm 1}}{\partial t} &=& M \left ( - \frac{\nu }{R_0^2} \widetilde{\delta R}_{\pm 1} + \eta_{\pm 1}(t) \right ),\\
\frac{\partial  \widetilde{\delta R}_{|m|\geqslant 2}}{\partial t} &=& M \left ( - \frac{\gamma}{R_0^2} (m^2-1) \widetilde{\delta R}_{m} + \eta_{m}(t) \right ).
\end{eqnarray}
These Langevin equations have the form $\dot x = -x/\tau+ M \eta(t)$, which solution is 
\beq
\label{sol_xt}
x(t) = x(0)e^{-t/\tau} + M \int \limits_0^t {e^{-(t -s)/\tau} \eta(s) ds}.
\eeq
The ensemble average results $\langle x(t) \rangle = \langle x(0) \rangle e^{-t/\tau}$, 
and $\langle x(t) \rangle \rightarrow 0$ for $t\rightarrow \infty$. 

The two time correlation function is 
\begin{equation}
\langle x(t)x^*(t') \rangle =  \langle |x(0)|^2 \rangle e^{ -(t+t')/\tau} + \frac{M^2 C\tau}{2}\left (  e^{ -|t-t'|/\tau} -  e^{ -(t+t')/\tau} \right ).
\label{ecn_Akttp}
\end{equation}
For $t=t'$, the static power spectrum is used in long time limit of 
\beq
\label{ecn_xt2}
\langle |x(t)|^2 \rangle =  \langle |x(0)|^2 \rangle e^{ -2 t/\tau} + \frac{M^2 C\tau}{2}\left (  1 -  e^{ -2 t/\tau} \right ),
\eeq
which gives the constant $C = 2\langle K_m \rangle/(\pi R_0 M)$. 

Finally, the mean square displacement is 
\beq
\label{msq_2}
\langle |\Delta R|^2\rangle = \langle |x(t) - x(0) |^2 \rangle.
\eeq
From Eqn. (\ref{sol_xt}) we obtain
\beq
x(t)-x(0) = x(0)(e^{-t/\tau}-1) + M \int \limits_0^t {e^{-(t -s)/\tau} \eta(s) ds},
\eeq
thus
\begin{eqnarray}
\langle |x(t)-x(0)|^2 \rangle &=& \langle | x(0)|^2 \rangle (e^{-t/\tau}-1)^2 + M^2 \int \limits_0^t \int \limits_0^t {e^{-(2t -s-s')/\tau} \langle \eta(s) \eta^*(s') \rangle  ds ds' },\\
& = & \langle | x(0)|^2 \rangle (e^{-t/\tau}-1)^2+ \frac{M^2 C\tau}{2}\left (  1 -  e^{ -2 t/\tau} \right ) \label{msd.temp},
 \end{eqnarray}
where we have used the fact that $\langle x(0) \eta(s) \rangle =  \langle x(0)\rangle \langle \eta(s) \rangle = 0$, for $s>0$, and the double integral is solved as before. We remind that the long time limit of (\ref{ecn_xt2}) gives $M^2C\tau/2 =  \langle | x(t)|^2 \rangle =  \langle | x(0)|^2 \rangle$, allowing to simplify expression (\ref{msd.temp}), leading to the following expression for the mean square displacement 
\beq
\langle |x(t)-x(0)|^2 \rangle =  M^2 C\tau\left (  1 -  e^{ -t/\tau} \right ).
\eeq
Finally, replacing the expression for $C$ in terms of the energy per mode,
\beq
\langle |x(t)-x(0)|^2 \rangle =  \frac{2 \langle K_m\rangle M\tau}{\pi R_0}\left (  1 -  e^{ -t/\tau} \right ).
\eeq

\subsection{Error analysis}

Both $|\widetilde {\delta R}_m|^2$ and $K_m$ have large fluctuations. Their distributions are strongly non-Gaussian, which implies that the corresponding error analysis has to be realized carefully. For example, the measured average kinetic energy for $m=0$ is $8.0$~nJ, whereas its standard deviation is $7.9$~nJ. Thus, the error related to this measurement is not its standard deviation. In fact, the associated error has to be computed with a generalized criterion. When a measurement of a quantity $x$ is performed, the usual procedure is to assign the standard deviation $\sigma_x$ as error to the average $\langle x \rangle$ of a set of data. For a Gaussian distribution, the range $[\langle x\rangle-\sigma_x,\langle x \rangle+\sigma_x$] corresponds to a confidence interval of $\approx 68\%$, meaning that if a new measurement is performed, it has a $68\%$ probability to be in this range. 

\begin{figure}[t!]
\begin{center}
\includegraphics[width=8.1cm]{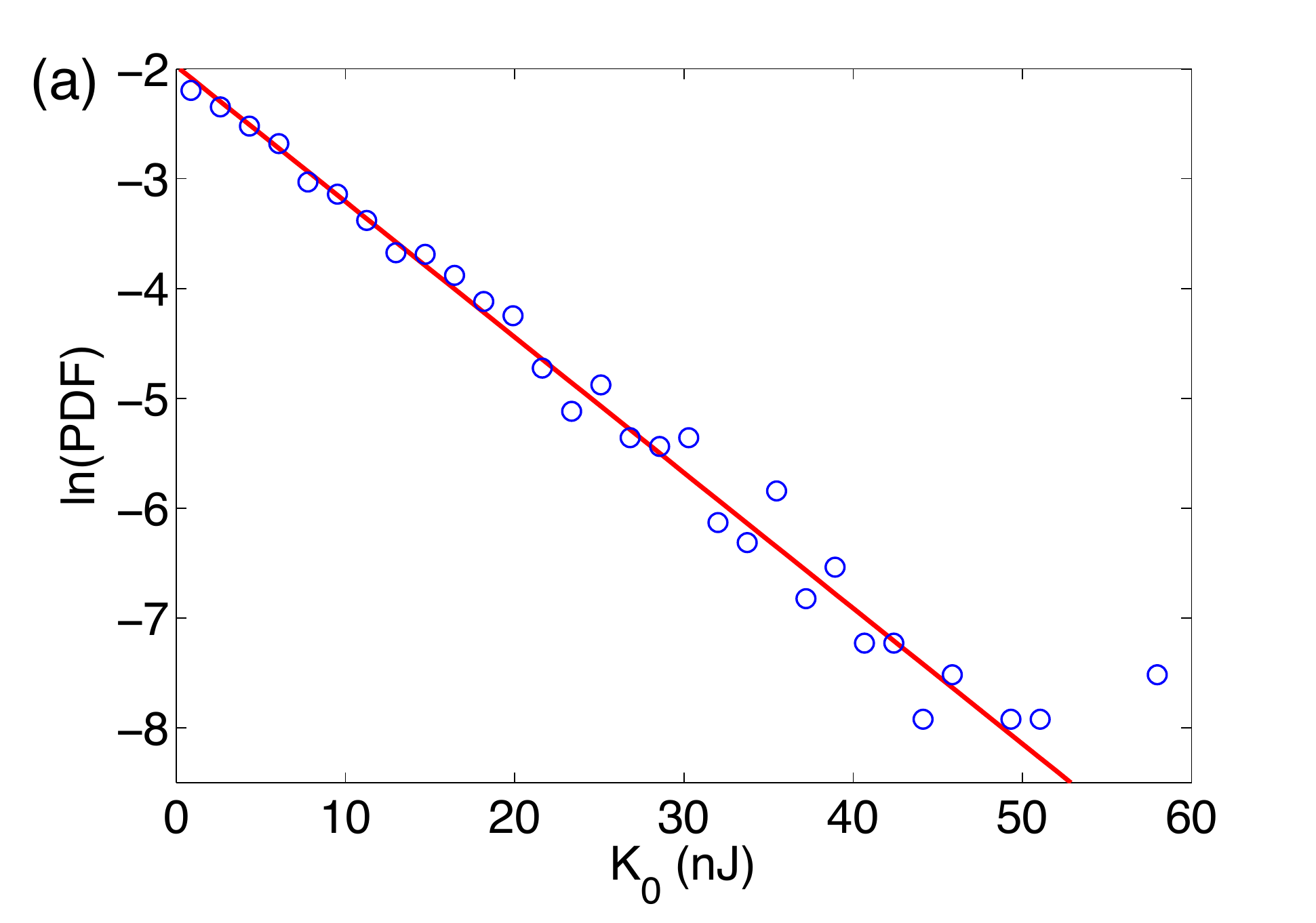}
\includegraphics[width=8.1cm]{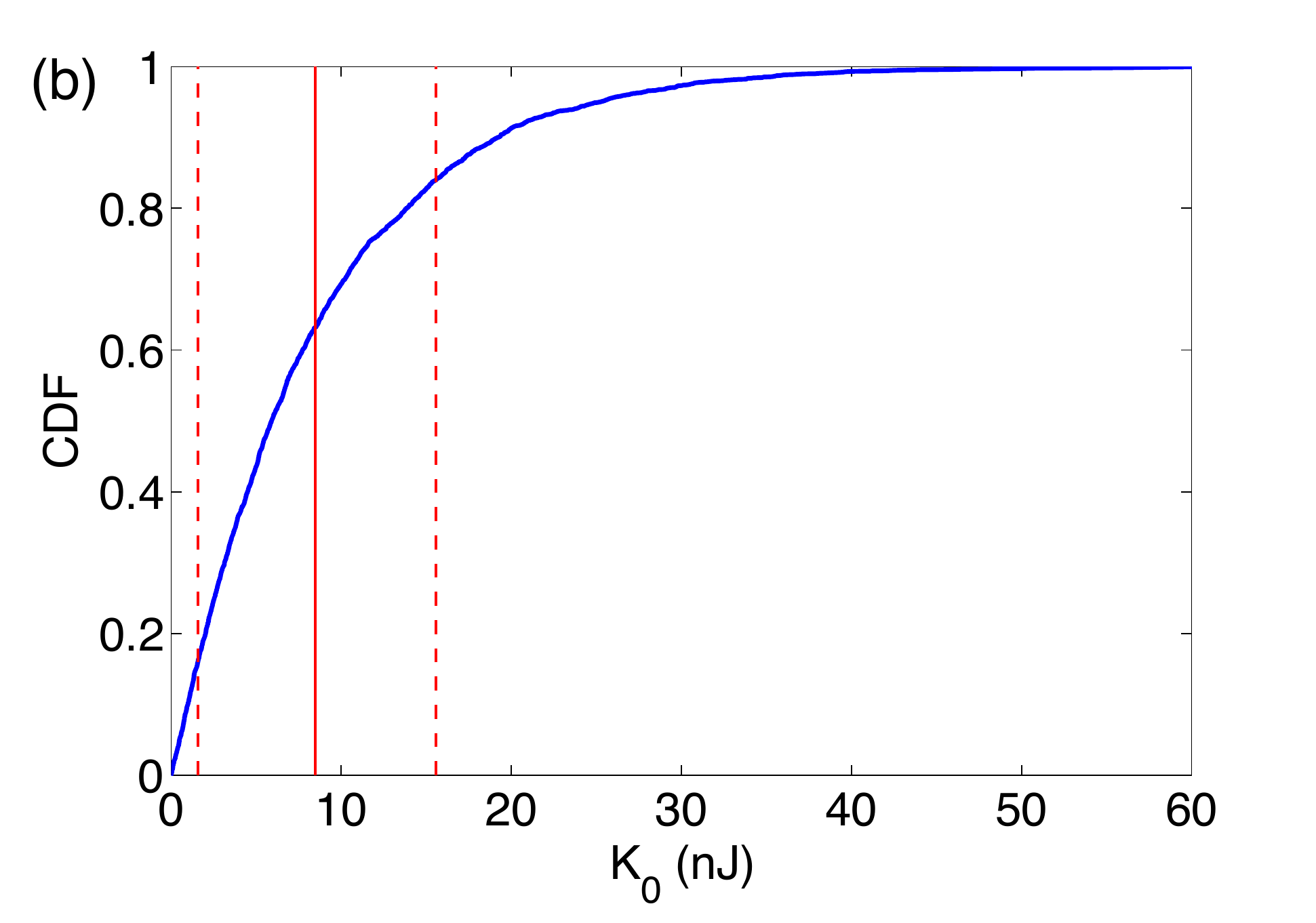}
\caption{(a) Probability density function, PDF, and (b) cumulative density function, CDF, of $K_0$ for one realization. In (a), the continuous line shows a Boltzmann distribution fit, ${\rm PDF}(K_0) = A \exp(-K_0/B)$, with $A=0.14\pm0.02$ (nJ)$^{-1}$ and $B = 8.1\pm0.4$ nJ. In (b), the vertical solid line corresponds to the average $\langle K_0 \rangle = 8.5$ nJ; the vertical dashed lines show the limits for which $16\%$ of the data is below $\langle K_0 \rangle-\sigma_{-}$, and $16\%$ is above $\langle K_0 \rangle+\sigma_{+}$, with $\sigma_{-} = 6.9$ nJ and $\sigma_{+} = 7.1$ nJ. For the other realization, $\langle K_0 \rangle = 7.5$ nJ, $\sigma_{-} = 6.2$ nJ and $\sigma_{+} = 6.1$ nJ.}
\label{fig1_supinfo}
\end{center}
\end{figure}

Figure \ref{fig1_supinfo}a shows that $K_0$ obeys a Boltzmann distribution, strongly non-Gaussian. Fig.~\ref{fig1_supinfo}b presents the cumulative distribution function (CDF) of the data presented in Fig. \ref{fig1_supinfo}a. The adopted criterium is the following: errors $\sigma_{-}$ and $\sigma_{+}$ are computed by estimating the values of the measured quantity that insure that $16\%$ of the data is lower that $\langle K_0 \rangle-\sigma_{-}$, and $16\%$ is larger that $\langle K_0 \rangle+\sigma_{+}$. Thus, $68\%$ of the data lies in the range $[\langle K_0 \rangle-\sigma_{-},\langle K_0 \rangle+\sigma_{+}]$. The same procedure is realized for each mode $m$. 

\begin{figure}[t!]
\begin{center}
\includegraphics[width=8.1cm]{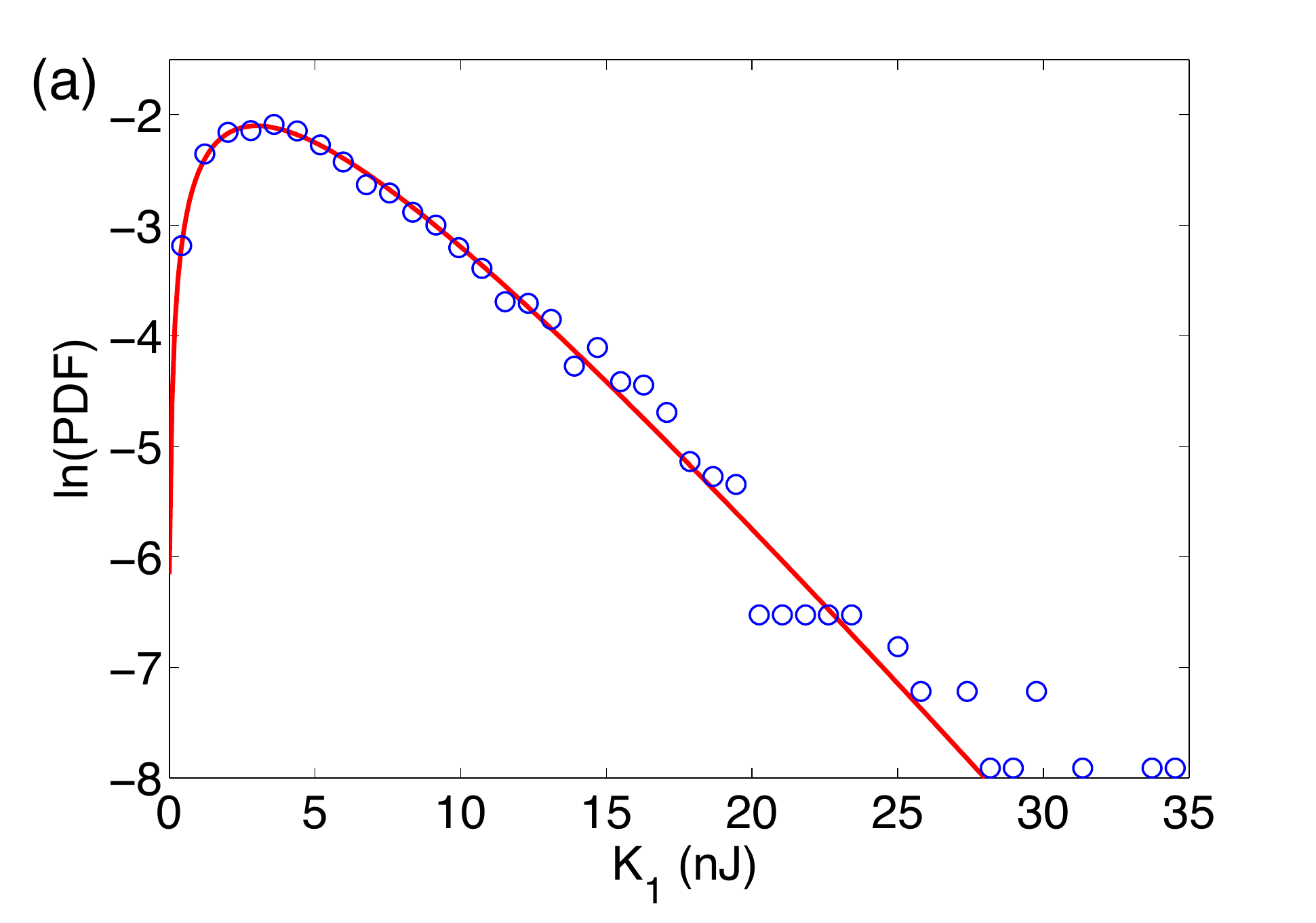}
\includegraphics[width=8.1cm]{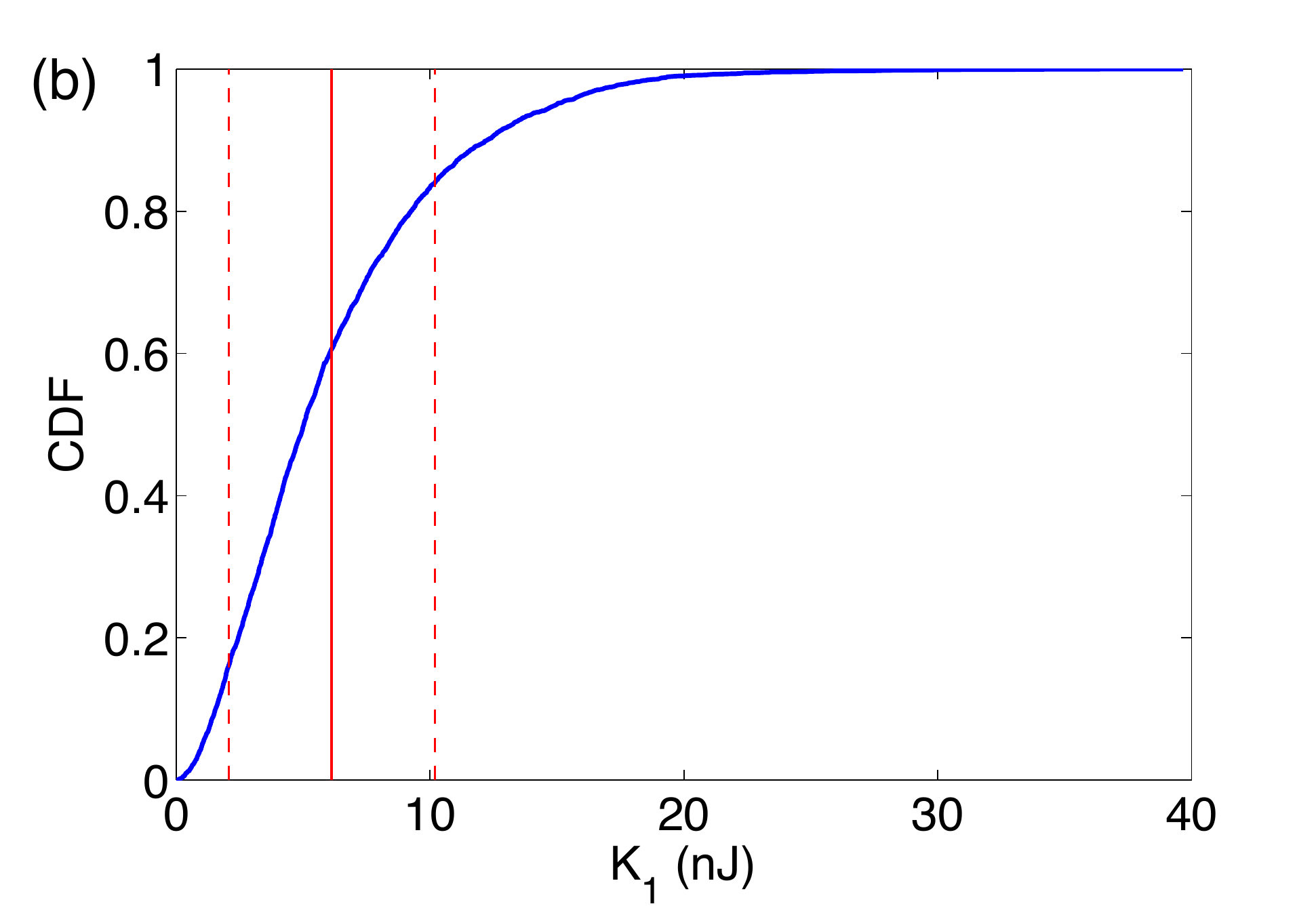}
\caption{(a) Probability density function, PDF, and (b) cumulative density function, CDF, of $K_1$ for one realization. In (a), the continuous line shows a generalized Poisson distribution fit, $\log [{\rm PDF}(K_1)/d] = -a(K_1+b) + c \log[a(K_1+b)]$, with $a=0.32\pm0.04$ (nJ)$^{-1}$, $b  = -0.02 \pm 0.60$ nJ, $c=1.0\pm0.5$ and $d = 0.33\pm0.06$ (nJ)$^{-1}$. In (b), the vertical solid line corresponds to $\langle K_1 \rangle = 6.1$~nJ; the vertical dashed lines show $\langle K_1 \rangle-\sigma_{-}$ and $\langle K_1 \rangle+\sigma_{+}$, with $\sigma_{-} = 4.1$ nJ and $\sigma_{+} = 4.1$ nJ. For the other realization, $\langle K_1 \rangle = 5.7$ nJ, $\sigma_{-} = 3.8$ nJ and $\sigma_{+} = 3.8$ nJ.}
\label{fig2_supinfo}
\end{center}
\end{figure}

Figure \ref{fig2_supinfo}a shows that $K_1$ obeys a generalized Poisson distribution, also strongly non-Gaussian. We have not yet a satisfactory explanation of why this fit seems to work very well, so it must be considered as a phenomenological fit. The distribution's asymmetry affects less the measured error bars. For the particular realization shown in this figure, $\sigma_{-} = 4.1$ nJ and $\sigma_{+} = 4.1$ nJ, with $\langle K_1 \rangle = 6.1$~nJ. However, the average is clearly larger than the most probable value $K_1^{\rm mp} = 3.0$ nJ (also known as the maximum likelihood value). For $m\geqslant2$, all $K_m$ obey the generalized Poisson distribution. Furthermore, between $m=2$ and $m\approx 30$ they all collapse on a single curve, which is consistent with the observed equipartition.

The fluctuations of $|\widetilde {\delta R}_m|^2$  also obey similar non-Gaussian, asymmetric distributions. Their error bars are computed using the same procedure described above. 

For both $|\widetilde {\delta R}_m|^2$ and $K_m$ the final asymmetric error bars are computed by averaging two independent realizations, thus reducing the total error by a factor $\approx\sqrt{2}$. This is what is known as the standard error for $N$ independent measurements, which scales as $\sigma/\sqrt{N}$, where $\sigma$ represents the error of each independent measurement. 
Because the asymmetries in the errors for both $|\widetilde {\delta R}_m|^2$ and $K_m$ are small, the reported error bars in the paper correspond to the maximum of $\sigma_{-}$ and $\sigma_{+}$, which allow an easier computation of errors when quantities with uncertainty are combined (sum, difference, multiplication or division) or when they are used for computing another quantity by means of a non-linear function.


\begin{thebibliography}{99}

\bibitem{urbach} P. Melby , F. Vega Reyes, A. Prevost, R. Robertson, P. Kumar, D. A. Egolf, and J. S. Urbach, J. Phys. Condens. Matter 17, S2689 (2005).

\bibitem{Castillo} G. Castillo, N. Mujica and R. Soto, Phys. Rev. Lett. {\bf 109}, 095701 (2012).

\end{thebibliography}
\end{document}